\documentclass[twocolumn,aps,prb,showpacs]{revtex4}
\usepackage{epsfig}

\begin{document}


\title{Dynamical spin response of electron doped Mott insulators on a triangular lattice}
\author{Bin Liu$^{1*}$ and Ying Liang$^{2}$}

\affiliation{$^{1}$Department of Physics, Beijing Jiaotong University, Beijing 100044, China\\$^{2}$Department of Physics, Beijing Normal University, Beijing 100875, China}

\begin{abstract}

The spin dynamics of electron doped Mott insulators on a triangular
lattice is studied based on the $t$-$J$ model. It is found that the
particularly universal behaviors of integrated dynamical spin
structure factor seen in the doped Mott insulators on a square
lattice, are absent in the doped Mott insulators on a triangular
lattice, indicating the presence of the normal state gap. As a
result, the spin-lattice relaxation rate 1/$T_{1}$ divided by $T$
reduces with decreasing temperatures in the temperature region above
0.2$J$$\approx$50K, then follows a Curie-Weiss-like behavior at the
temperature less than 50K, in qualitative agreement with
experimental observations.

\end{abstract}
\pacs{71.70.Ej, 73.20.At, 74.20.-z}

\maketitle


\section{ Introduction}
In the past few years, the discovery of superconductivity in layered
cobalt oxides Na$_{x}$CoO$_{2}\cdot y$H$_{2}$O\cite{takada} has
attracted considerable attention among the condensed matter physics
society. Although the superconducting transition temperature
($T_{c}\approx 4.5$K)\cite{takada} is very low, the layered
electronic structure and the dome-shaped behavior of $T_{c}$ suggest
this hydrated compound may be another doped Mott
insulator\cite{takada,schaak}. In contrast to the cuprate
superconductors with a square lattice of CuO$_{2}$ planes, the
parent compound Na$_{x}$CoO$_{2}$ is composed of two-dimensional
(2D) CoO$_{2}$ layer with triangular geometry. This geometry
frustration results in some novel electronic and magnetic phases,
for instance, the Anderson's resonating valence state
\cite{anderson0} and the strong topological frustration
phases\cite{baskaran,liu}. With the doping variation,
Na$_{x}$CoO$_{2}$ displays rich and complicated phase
diagram\cite{foo}. In particular, superconductivity appears around
$x=0.3$ when water molecules are intercalated between the CoO$_{2}$
layers\cite{takada,schaak}.

Although extensive researches on this compound have been done in
both experimental and theoretical sides, the doping dependent
magnetic properties still remain unclear. The local spin density
approach\cite{sing} predicted ferromagnetic coupling within 2D
CoO$_{2}$ layer for nearly all electron doping $0.3\leq x \leq0.7$.
In fact, up to now, only at large doping neutron scattering studies
and time-of-flight experiments on Na$_{0.75}$CoO$_{2}$ and
Na$_{0.82}$CoO$_{2}$\cite{boothroyd,bayrakci} respectively, present
low-energy fluctuations characteristic of A-type antiferromagnetism:
antiferromagnetically coupled ferromagnetic layers. However, with
the decreasing of Na content, a series of nuclear magnetic resonance
(NMR)\cite{yokoi} and nuclear quadrupole resonance
(NQR)\cite{fujimoto} as well as Knight shift measurement\cite{zheng}
show the presence of strong 2D antiferromagnetic (AF) spin
correlations and suggest the possibility of the spin-singlet
superconductivity. That is to say the 2D AF spin correlations may be
realized in the region of small $x$ value for Na$_{x}$CoO$_{2}$.
Starting from ab-inito band structure calculations, Korshunov and
co-workers have estimated a certain critical concentration $x_m$,
below which the magnetic susceptibility within the CoO$_{2}$ plane
shows a tendency towards AF fluctuations\cite{korshunov}. The recent
correlation effects on the doped Na$_{x}$CoO$_{2}$ via a cellular
cluster approach has reconfirmed and explained such in-plane
magnetic transition from FM tendencies towards to AF\cite{frank}.

Recently, the surface-sensitive angle resolved photo emission
spectroscopy (ARPES) experiments\cite{hasan,yang1,yang2,qian}
observe a doping-dependent evolution of the fermi surface, which is
found to be centered around the $\Gamma$ point with a mostly
$a_{1g}$ characterister and no sign of six pockets resulting from
the $e^{'}_{g}$ band predicted by local density approximation
calculation\cite{sing} for a wide of Na concentrations. This
indicates the importance of the electronic corrections in the doped
Na$_{x}$CoO$_{2}$. Therefore, as discussed by many
researchers\cite{baskaran,liu,kumar,wang}, an effective single band
low-energy model suffices to capture the essential physics of
electron-doped cobaltate Na$_{x}$CoO$_{2}$. With this consideration
in mind, in this paper we also take a single band $t$-$J$ model as
our starting point to discuss its unusual dynamical spin response.
Since the AF fluctuations within the CoO$_{2}$ planes may be
realized in the low electron doping range, we are focusing on such
region in our calculations below. We find that  the particularly
universal behaviors of integrated dynamical spin structure factor
seen in the doped cuprates, are absent in the doped cobaltates,
indicating the presence of the normal state gap. We also find that
the spin-lattice relaxation rate 1/$T_{1}$ divided by $T$ reduces
with decreasing temperatures in the temperature region above
0.2$J$$\approx$50K, then follows a Curie-Weiss-like behavior at the
temperature less than 50K originating from the strong AF
fluctuations, in qualitative agreement with experimental
observations\cite{yokoi, fujimoto}. Our results also show that AF
fluctuations within the CoO$_{2}$ plane gradually weaken with
increasing electron doping.

\section{the $t$-$J$ Model and Fermion-spin theory}

It has been argued that the essential physics of the doped CoO$_{2}$
plane is contained in the $t$-$J$ model on a triangular lattice\cite{baskaran},
\begin{eqnarray}
H&=&-t_{e}\sum_{i\hat{\eta}\sigma}PC_{i\sigma}^{\dagger}
C_{i+\hat{\eta}\sigma}P^{\dagger}-\mu\sum_{i\sigma}P C_{i\sigma
}^{\dagger}C_{i\sigma }P^{\dagger}\nonumber\\&+&
J\sum_{i\hat{\eta}}{\bf S}_{i}\cdot{\bf S}_{i+\hat{\eta}},
\end{eqnarray}
where the summation is over all sites $i$, and for each $i$, over
its nearest-neighbor $\hat{\eta}$, $C^{\dagger}_{i\sigma}$
($C_{i\sigma}$) is the electron creation (annihilation) operator,
${\bf S}_{i}=C^{\dagger}_{i}{\bf \sigma}C_{i}/2$ is the spin
operator with ${\bf \sigma}=(\sigma_{x},\sigma_{y},\sigma_{z})$ as
the Pauli matrices, $\mu$ is the chemical potential, and the
projection operator $P$ removes zero occupancy, i.e.,
$\sum_{\sigma}C^{\dagger}_{i\sigma} C_{i\sigma}\geq 1$. In the past
fifteen years, some useful methods have been proposed to treat the
no double occupancy local constraint in hole doped cuprates. In
particular, a fermion-spin theory based on the partial charge-spin
separation has been developed to study the physical properties of
doped cuprates\cite{feng1,feng2}, where the no double occupancy local
constraint can be treated properly in analytical calculations. To
apply this theory in the electron doped cobaltates, the $t$-$J$
model (1) can be rewritten in terms of a particle-hole
transformation $C_{i\sigma}\rightarrow f^{\dagger}_{i-\sigma}$ as,
\begin{eqnarray}
H=-t\sum_{i\hat{\eta}\sigma}f_{i\sigma}^{\dagger}
f_{i+\hat{\eta}\sigma}+\mu\sum_{i\sigma}f_{i\sigma }^{\dagger}
f_{i\sigma }+J\sum_{i\hat{\eta}}{\bf S}_{i} \cdot{\bf
S}_{i+\hat{\eta}},
\end{eqnarray}
supplemented by the local constraint
$\sum_{\sigma}f^{\dagger}_{i\sigma}f_{i\sigma}\leq 1$ to remove
double occupancy, where for convenience, we have set $t=-t_{e}<0$,
$f^{\dagger}_{i\sigma}$ ($f_{i\sigma}$) is the hole creation
(annihilation) operator. Then the hole operators can be decoupled
as, $f_{i\uparrow}=a^{\dagger}_{i\uparrow}S^{-}_{i}$ and
$f_{i\downarrow}=a^{\dagger}_{i\downarrow}S^{+}_{i}$, in the
charge-spin separation fermion-spin theory\cite{feng1,feng2}, where the
spinful fermion operator $a_{i\sigma}=e^{-i\Phi_{i\sigma}}a_{i}$
describes the charge degree of freedom together with some effects of
the spin configuration rearrangements due to the presence of the
doped electron itself (charge carrier), while the spin operator
$S_{i}$ describes the spin degree of freedom (spin), then the single
occupancy local constraint, $\sum_{\sigma}
f^{\dagger}_{i\sigma}f_{i\sigma} =S^{+}_{i}a_{i\uparrow}
a^{\dagger}_{i\uparrow} S^{-}_{i}+ S^{-}_{i}a_{i\downarrow}
a^{\dagger}_{i\downarrow} S^{+}_{i}= a_{i}a^{\dagger}_{i} (S^{+}_{i}
S^{-}_{i}+S^{-}_{i} S^{+}_{i})=1- a^{\dagger}_{i} a_{i}\leq 1$, is
satisfied in analytical calculations. In this charge-spin separation
fermion-spin representation, the low-energy behavior of the $t$-$J$
model (2) can be expressed as\cite{liu1},
\begin{eqnarray}
H&=&-t\sum_{i\hat{\eta}}(a_{i\uparrow}S^{+}_{i}
a^{\dagger}_{i+\hat{\eta}\uparrow}S^{-}_{i+\hat{\eta}}+
a_{i\downarrow}S^{-}_{i}a^{\dagger}_{i+\hat{\eta}\downarrow}
S^{+}_{i+\hat{\eta}})\nonumber\\&-&\mu\sum_{i\sigma}a^{\dagger}_{i\sigma}
a_{i\sigma}+J_{{\rm eff}}\sum_{i\hat{\eta}}{\bf S}_{i}\cdot {\bf
S}_{i+\hat{\eta}},
\end{eqnarray}
with $J_{{\rm eff}}=(1-x)^{2}J$, and $x=\langle
a^{\dagger}_{i\sigma}a_{i\sigma}\rangle=\langle a^{\dagger}_{i}
a_{i}\rangle$ is the electron doping concentration. In this case,
the magnetic energy ($J$) term in the $t$-$J$ model is only to form
an adequate spin configuration, while the kinetic energy ($t$) term
has been transferred to the charge carrier-spin interaction, which
dominates the essential physics.

In the framework of the charge-spin separation, the basic low-energy
excitations are charge carriers and spins. It has been shown that
the charge dynamics is mainly governed by the scattering from the
charge carriers due to the spin fluctuation\cite{liu1}, while the
spin fluctuations couple only to the spin and therefore no
composition law is required in discussing the spin
dynamics\cite{feng5}, but the effect of charge carriers is still
considered through the charge carrier's order parameter $\phi$
entering the spin propagator. In this case, the spin dynamics of the
doped square antiferromagnet has been discussed\cite{feng5} by
considering the spin fluctuation around the mean-field solution,
where the spin part is treated by the loop expansion to the second
order. Following their discussions, we can obtain the dynamical spin
structure for the electron doped cobaltates as
\begin{eqnarray}
S(k,\omega)&=&{\rm Re}\int_{0}^{\infty}dte^{iw(t-t')}D(k,t-t')\nonumber\\
&=&-2[1+n_{B}(\omega)]{\rm Im}D(k,\omega),
\end{eqnarray}
respectively, where the full spin Green's function,
$D^{-1}(k,\omega)=D^{(0)-1}(k,\omega)-\Sigma_{s}(k,\omega)$, with
the mean-field (MF) spin Green's functions
$D^{(0)-1}(k,\omega)=(\omega^{2}-\omega_{k}^{2})/B_{k}$, and the
second-order spin self-energies from the charge carrier pair bubble,
\begin{eqnarray}
&~&\Sigma_{s}(k,\omega)=\left ({Zt\over N}\right )^{2}
\sum_{pp'}(\gamma^{2}_{p'+p+k}+\gamma^{2}_{p'-k})
{B_{k+p}\over 2\omega_{k+p}}\nonumber\\
&\times&({F^{(s)}_{1}(k,p,p')\over\omega+\xi_{p+p'}-\xi_{p'}
-\omega_{k+p}}-{F^{(s)}_{2}(k,p,p')\over \omega+\xi_{p+p'}-\xi_{p'}
+\omega_{k+p}}),
\end{eqnarray}
where $Z$ is the number of the nearest neighbor sites,
$B_{k}=\lambda[2\chi^{z}(\epsilon
\gamma_{k}-1)+\chi(\gamma_{k}-\epsilon)]$, $\lambda=2ZJ_{{\rm
eff}}$, $\epsilon=1+2t\phi/J_{{\rm eff}}$, $\gamma_{k}=[\cos{k_{x}}+
2\cos{(k_{x}/2)}\cos{({\sqrt 3}k_{y}/2)}]/3$, the spin correlation
function $\chi^{z}=\langle S_{i}^{z}S_{i+\hat{\eta}}^{z}\rangle$,
$F^{(s)}_{1}(k,p,p')=n_{F}(\xi_{p+p'})[1-n_{F}(\xi_{p'})]-n_{B}
(\omega_{k+p})[n_{F}(\xi_{p'})-n_{F}(\xi_{p+p'})]$, $F^{(s)}_{2}
(k,p,p')=n_{F}(\xi_{p+p'})[1-n_{F}(\xi_{p'})]+[1+n_{B}
(\omega_{k+p})][n_{F}(\xi_{p'})-n_{F}(\xi_{p+p'})]$,
$n_{B}(\omega_{p})$ and $n_{F}(\xi_{p})$ are the boson and fermion
distribution functions, respectively. The MF charge carrier and spin
spectra are given by $\xi_{k}=Zt\chi\gamma_{k}-\mu$, and
$\omega^{2}_{k}=A_{1}(\gamma_{k})^{2}+A_{2}\gamma_{k}+A_{3}$,
respectively, with
$A_{1}=\alpha\epsilon\lambda^{2}(\epsilon\chi^{z}+\chi/2)$,
$A_{2}=-\epsilon\lambda^{2}[\alpha(\chi^{z}+\epsilon\chi/2)+ (\alpha
C^{z}+(1-\alpha)/(4Z)-\alpha\epsilon\chi/(2Z))+(\alpha
C+(1-\alpha)/(2Z)-\alpha\chi^{z}/2)/2]$, $A_{3}=\lambda^{2} [\alpha
C^{z}+(1-\alpha)/(4Z)-\alpha\epsilon\chi/(2Z)+\epsilon^{2} (\alpha
C+(1-\alpha)/(2Z)-\alpha\chi^{z}/2)/2]$, and the spin correlation
function $\chi=\langle S_{i}^{+}S_{i+\hat{\eta}}^{-} \rangle$,
$C=(1/Z^{2})\sum_{\hat{\eta}\hat{\eta^{\prime}}}\langle
S_{i+\hat{\eta}}^{+}S_{i+\hat{\eta^{\prime}}}^{-}\rangle$, and
$C^{z}=(1/Z^{2})\sum_{\hat{\eta},\hat{\eta'}}\langle
S_{i+\hat{\eta}}^{z}S_{i+\hat{\eta'}}^{z}\rangle$. In order not to
violate the sum rule of the correlation function $\langle
S^{+}_{i}S^{-}_{i}\rangle=1/2$ in the case without AF long-range
order, the important decoupling parameter $\alpha$ has been
introduced in the MF calculation\cite{kondo,feng4}, which can be
regarded as a vertex correction. All the above MF order parameters,
decoupling parameter $\alpha$, and chemical potential $\mu$ are
determined by the self-consistent calculation\cite{feng4}.

\section{Numerical Results and Discussions}

\begin{figure}[ht]
\epsfxsize=3.5in\centerline{\epsffile{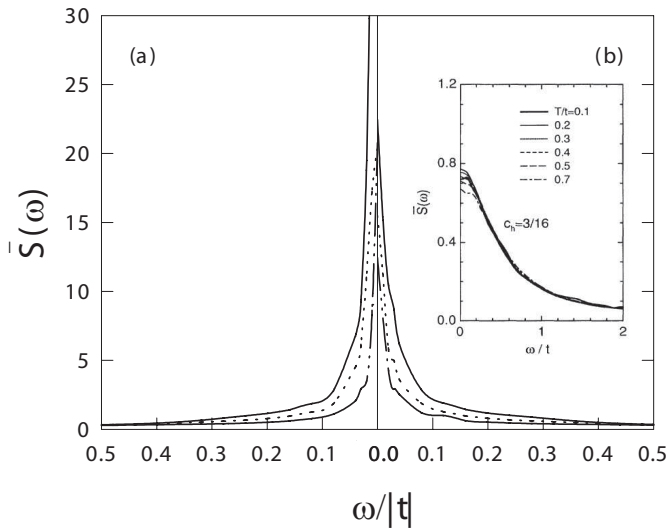}} \caption{The
integrated dynamical spin structure factor at the doping (a)
$x=0.33$ and (b) $x=0.30$ with the temperature $T=0.3J$
(dashed-dotted line), $T=0.4J$ (dotted line), and $T=0.6J$ (solid
line) for the parameter $t/J=-2.5$. Inset: the
numerical simulation taken from Ref. [27].}
\end{figure}
\begin{figure}[ht]
\epsfxsize=3.5in\centerline{\epsffile{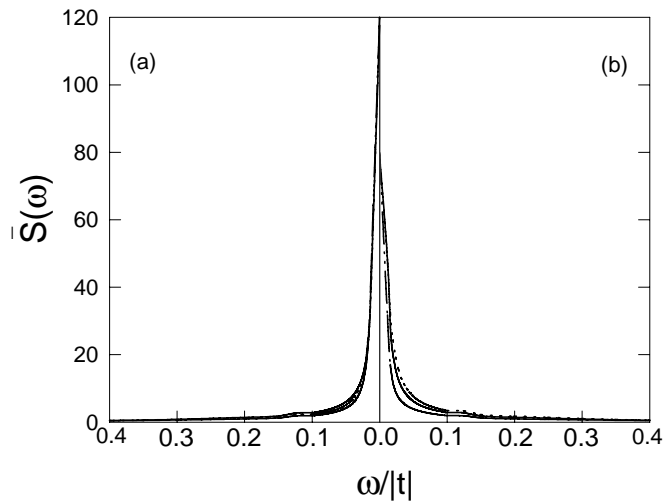}} \caption{The
integrated dynamical spin structure factor at the doping (a)
$x=0.33$ and (b) $x=0.30$ with the temperature $T=0.7J$
(dashed-dotted line), $T=0.8J$ (dotted line), and $T=0.9J$ (solid
line) for the parameter $t/J=-2.5$.}
\end{figure}

For understanding the dynamic spin response of the electron doped
cobaltates, we have calculated the integrated dynamical spin
structure factor, which can be expressed as
\begin{eqnarray}
\overline{S}(\omega,T)&=&S_{L}(\omega)+S_{L}(-\omega)=(1+e^{-\beta\omega})S_{L}(\omega)\nonumber\\
&=&(1+e^{-\beta\omega})\frac{1}{N}\sum_{k}S(k,\omega).
\end{eqnarray}
The numerical results of the integrated dynamical spin structure
factor at doping (a) $x=0.33$ and (b) $x=0.30$ with the temperature
$T=0.3J$ (dashed-dotted line), $T=0.4J$ (dotted line), and $T=0.6J$
(solid line) for the parameter t/J= -2.5 are plotted in Fig. 1. It
is shown that the integrated dynamical spin structure factor
decreases with increasing energies for $\omega < 0.3t$, and almost
constant for $\omega \geq 0.3t$. which is similar to the case of
doped cuprates\cite{keimer,feng5} and reflects the common features
of the doped Mott insulators. However, the temperature dependence of
the integrated dynamical spin structure factor, which is essentially
temperature independent at high energy in a wide temperature regime,
presents a strong deviation at the low-energy range, so that the
shape of the integrated dynamical spin structure factor does not
follow a universal behavior, in contrast to the case of the doped
cuprates (seen the insert in Fig. 1b)\cite{keimer}. It has been discussed that the universal
behavior of the integrated dynamical spin structure factor in the
doped cuprates is due to the absence of the normal state gap in the
electron density of states\cite{feng5}. However, in the doped
cobaltates, the normal state gap opens\cite{yokoi,yada,feng6},
which is mainly induced by the frustrated spin. Since in the charge-spin separation
framework the spin dynamics is dominated by the scattering of spins,
which are strongly renormalized because of the strong interactions
with fluctuations of surrounding charge carrier excitations. The
frustrated spin moves in the background of the charge carriers, and
the cloud of distorted charge carriers background is to follow the
frustrated spins, which leads to the anomalous spin dynamics in the
doped cobaltates. In other words, the origin of the absence of the
particularly universal behavior of the integrated dynamical spin
structure factor in the doped cobaltates at low energy originates from
the normal-state gap due to the magnetic frustration. In order to show this normal state
gap clearly, we calculate the $\overline{S}(\omega,T)$ at higher
temperature as shown in Fig. 2, where the parameters are the same
with the Fig. 1 except for the temperature $T=0.7J$ (dashed-dotted
line), $T=0.8J$ (dotted line), and $T=0.9J$ (solid line). In Fig. 2,
the temperature dependent integrated dynamical spin structure factor
nearly degenerates and doesn't show a deviation at low energy seen
in Fig. 1. This universal behavior of the integrated dynamical spin
structure factor exists for the whole temperature range is similar
to the case of doped cuprates (seen the insert in Fig. 1b)\cite{keimer,feng5}. From Fig. 1 and
Fig. 2, we estimate the normal state gap opens at temperature
$T_{gap}\simeq0.7J$. According to our estimation
$J\simeq250K$\cite{liu}, we get the $T_{gap}\simeq160K$, very close
to the experiments\cite{yokoi}.

Now we turn our attention to discuss the spin-lattice relaxation
rate 1/$T_{1}$. Theoretically, the spin contributions to the
spin-lattice relaxation rate 1/$T_{1}$ divided by $T$ may be written
using the imaginary part of the Co dynamical electron
spin-susceptibility $\chi''({\bf k},\omega)$ as
\begin{eqnarray}
{1\over T_{1}T}&=&{2\gamma^{2}_{i}k_{B}\over
g^{2}\mu_{B}^{2}}\lim_{\omega\rightarrow0}\sum_{k}|A(k)|^{2}{\chi''({\bf
k},\omega)\over \omega} ,
\end{eqnarray}
where $\chi''(k,\omega)=[1+n_{B}(\omega)]^{-1}S(k,\omega)$, $g$ is
the $g$ factor, $\mu_{B}$ is the Bohr magneton, and
$A(k)=\sum_{i}A_{i}e^{ikr_{i}}$, where $A_{i}$ is the hyperfine
coupling between the electron spin and the nuclear spin having
gyromagnetic ratio $\gamma_{i}$\cite{shastry} and is to be
approximated to a constant in our calculation.
\begin{figure}[ht]
\epsfxsize=3.5in\centerline{\epsffile{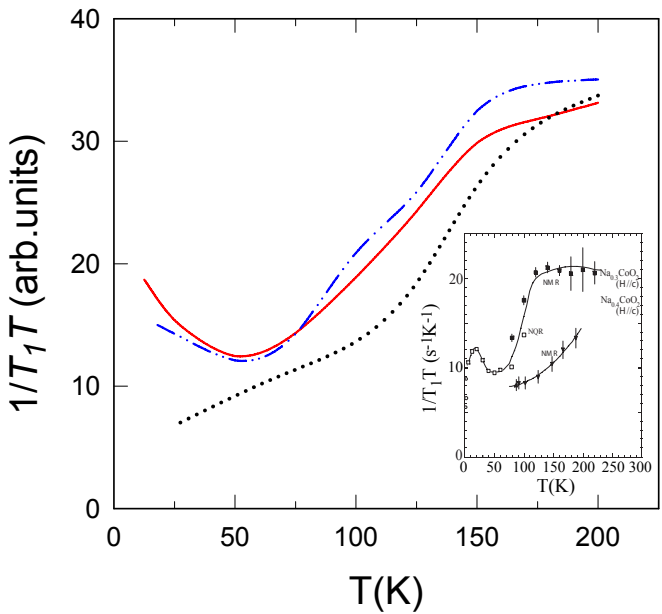}} \caption{(color online) The
spin-lattice relaxation rate 1/$T_{1}$ divided by $T$ of the
electron doped cobaltates as a function of temperature at doping
$x=0.30$ (solid line), $x=0.33$ (dashed-dotted line), and $x=0.40$
(dotted line) for parameter $t/J=-2.5$, where we choose
${2\gamma^{2}_{i}k_{B}A_{i}/ g^{2}\mu_{B}^{2}}$=constant. Inset: the
experimental result taken from Ref. [10].}
\end{figure}The spin-lattice
relaxation rate 1/$T_{1}$ divided by $T$ in Eq.(7) has been
evaluated numerically and the results for t/J= -2.5 at the doping
x=0.30 (solid line), x=0.33 (dashed-dotted line), and x=0.40 (dotted
line) in comparison with the corresponding experiments (inset) are
plotted in Fig. 3. It is shown that at low doping x=0.30 (solid
line) and x=0.33 (dashed-dotted line), the spin-lattice relaxation
rate 1/$T_{1}T$ reduces with decreasing temperature in the
temperature region $0.2J\approx50K < T < 0.7J\approx160K$, which is
self-consistent with the integrated dynamical spin susceptibility
shown in Fig. 1 and Fig. 2, and is the result of the opening of
normal state gap. However, in the low temperature region $ T <
0.2J$, the spin-lattice relaxation rate 1/$T_{1}T$ follows a
Curie-Weiss-like behavior due to the existence of strong AF
fluctions. However, at the doping x=0.40 (dotted line), although the
normal state gap still exits at high temperature, Curie-Weiss-like
behavior at low temperature disappears, indicating that
AF-fluctuations have become weak with the increasing doping. These
results are in good agreement with the recent
experiments\cite{fujimoto,yokoi}. At the high doping $x > 0.4$, the
neutron scattering studies and time-of-flight experiments on
Na$_{0.75}$CoO$_{2}$ and
Na$_{0.82}$CoO$_{2}$\cite{boothroyd,bayrakci}, respectively, have
observed the ferromagnetic correlation in the 2D CoO$_{2}$ plane.
Therefore, the magnetic fluctuation in the 2D CoO$_{2}$ plane has
been affected by the electron doping, $i.e.$, the magnetic
fluctuation is doping dependent. The later ab-inito band structure
calculations for Na$_{x}$CoO$_{2}$ have given a certain critical
value $x_m=0.56\sim0.68$. The magnetic susceptibility in the
CoO$_{2}$ plane shows a tendency towards the AF state below $x_m$
and ferromagnetic state above $x_m$\cite{korshunov}.

\section{Summary}

In summary, we have studied the dynamical spin response in the
electron doped cobaltates within the $t$-$J$ model. We have shown
that due to the presence of normal state gap, the particularly
universal behaviors of integrated dynamical spin structure factor
seen in the doped cuprates at the low energy, are absent in the
doped Mott insulators on a triangular lattice. We have also shown
that the spin-lattice relaxation rate 1/$T_{1}$ divided by $T$
reduces with decreasing temperature in the temperature region above
0.2$J$$\approx$50K, then follows a Curie-Weiss-like behavior at the
temperature less than 50K originating from the strong
antiferromagnetic fluctuations, in qualitative agreement with
experiments. Our results also indicate that the electron doping
plays a important role in altering the magnetic properties of the
CoO$_{2}$ plane.

\section{acknowledgments}
This work was supported by the National Natural Science Foundation of China (NSFC) under Grants No. 11104011, Fundamental Research Funds for the Central Universities in China under Grant No. 2011JBM126, Research Funds of Beijing Jiaotong University under Grant No. 2011RC027, and National
Basic Research Program of China under Grant No.
2011CBA00102. We thank helpful discussions with Shiping Feng, I.
Eremin, M.M. Korshunov, and Guo-qing Zheng.

\end{document}